\documentclass[a4paper]{article}
\usepackage{amsfonts}
\usepackage{latexsym}
\usepackage{amsmath}
\usepackage{mathrsfs}

\begin{document}
\title{Description of all conformally invariant differential operators,
 acting on scalar functions}
\author{Petko Nikolov\,\dag\, , Tihomir Valchev\,\ddag \\
\small\dag\,Department of Theoretical Physics, Sofia University,\\
\small 5, James Bourchier Blvd., 1164 Sofia, Bulgaria \\
\small\ddag\,Institute for Nuclear Research and Nuclear Energy,Bulgarian Academy of \\
\small Sciences, 72 Tzarigradsko chausse\`{e}, 1784 Sofia,
Bulgaria}
\date{}
\maketitle

\begin{abstract}
We give an algorithm write down all conformally invariant
differential operators acting between scalar functions on
Minkowski space. All operators of order $k$ are nonlinear and are
functions on a finite family of functionally independent invariant
operators of order up to  $k$. The independent differential
operators of second order are three and we give an explicit
realization of them. The applied technique is based on the jet
bundle formalism, algebraization of the the differential
operators, group action and dimensional reduction. As an
illustration of this method we consider the simpler case of
differential operators between analytic functions invariant under
the modular group. We give a power series generating explicitly
all the functionally independent invariant operators of an
arbitrary order.
\end{abstract}

\section{Introduction}
It is well-known that the Maxwell equations are conformally
invariant. This motivate an permanent interest in studying the
conformal classes of metrics, conformally invariant operators and
structures. The Maxwell differential operator is linear and so it
is a splitting operator between two (infinitely dimensional)
linear representation of the conformal group. There are many
papers in the literature treating the splitting operators between
different representations of the conformal group. These operators
generalize the Maxwell operator with respect of this property of
invariance. These considerations are based on the studying of the
description and the structure of the linear representations and
their subrepresentations.

The aproach applied in the present work is different. It is based
on jet bundles technique (\cite{jet1}), (\cite{jet3}) and
(\cite{jet4}). The jet lifting of functions (or sections) plays a
role of an universal differential operator. The differential
operators are viewed as a composition of a jet lifting and a fibre
preserving map (a morphism for the case of linear operators)
between vector bundles. This algebraise the differential operators
and after this the invariance means invariance of these maps
(morphisms). During the next step the technique of group action
and dimensional reduction are used to describe the invariant
fibre-preserving maps (sections). A crucial point is the study of
the action of stationary subgroup of some point on the jets of
smooth sections at this point. This contains elements of the
Catastrophe theory in the sense of R. Tom and Arnold (see
\cite{catastrophes1}and \cite{catastrophes2}). If the steps have
been done explicitly this prescription gives all invariant
operators including the nonlinear ones.

In our case (differential operators acting on scalar functions on
Minkowski space) the main result is the following: all invariant
operators are nonlinear of order equal or bigger than two (we
exclude the trivial case of zero order operators). For second
order ($n=2$) there are three independent differential operators
(see \ref{D1}, \ref{D2} and \ref{D3}). Any other invariant
operator of order two is a function of three variables of them.
ASimilarly, for order $n=3$ there are $23$ independent
operators---the previous three and $20$ operators more of order
three. Every invariant operator up to order three is a function of
these universal operators. For the case of an arbitrary finite
order $n$ there exist a finite complete system of independent
conformally invariant operators up to order $n$ i.e. they generate
all other invariant operators up to the corresponding order.

In the simpler case of differential operators between analytic
functions, invariant under the modular group the result is
similar. All invariant operators are nonlinear.  There are $n-2$
independent operators of the order up to $n$. All others of are
functions of them. We give a power series generating the family of
independent invariant differential operators. The invariant
operator of order three is closely related to the Schwarz
derivative (there is a difference because the Schwarz derivative
takes values in the quadratic differentials).

\section{General Scheme}

Let $\xi=(E,p.M)$ be a vector bundle where \emph{E} is a
$m+n$-dimensional manifold and \emph{M} is a \emph{m}-dimensional
manifold. Smooth sections $M\to E$ build an infinite dimensional
space denoted by $C^{\infty}(\xi)$. We will often use adapted
coordinates $(x^{\mu},u^a)$, i.e. coordinates satisfying the
relation $p(x^{\mu},u^a)=(x^{\mu})$, $\mu=1,\ldots ,m,\ a=1,\ldots
,n$.

A connected Lie group $G$ acts on $\xi$ by bundle morphisms
\begin{displaymath}
\forall\ \ g\in G\ \exists\ T_{g}\in Diff(E):\
T_{g}:\xi_{x}\rightarrow\xi_{t_{g}(x)}\ ,
\end{displaymath}
where $\xi_{x}=p^{-1}(x)$ is the fibre over $x\in M$ and $t_g\in\,
Diff(M)$ is the projection of a morphism $T_g$. In adapted
coordinates $(x^{\mu},u^a)$ the action of $G$ reads
\begin{displaymath}
(x^\mu,u^a)\to
\left((t_g)^\mu(x),\left(T_g\right)^a_b(x)u^b(x)\right).
\end{displaymath}
The group $G$ has a natural action on $C^{\infty}(\xi)$
\begin{displaymath}
G\times C^{\infty}(\xi)\ni(g,\psi)\longrightarrow g(\psi)\in
C^{\infty}(\xi).
\end{displaymath}
\begin{equation}
g(\psi)(x):=T_{g}\left(\psi\left(t^{-1}_{g}(x)\right)\right).
\end{equation}
In the local coordinates $(x^{\mu},u^{a})$ this action is given by
\begin{displaymath}
g(\psi)^{a}(x)=\left(T_{g}\right)^{a}_{b}\left(t^{-1}_{g}(x)\right)\psi^{b}
\left(t^{-1}_{g}(x)\right).
\end{displaymath}
A natural problem is the description of the vector subspace
$C^{\infty}(\xi)_{G}\subset C^{\infty}(\xi)$ of all
\emph{G}-invariant sections. The invariance condition
$g(\psi)=\psi$ locally looks as
\begin{displaymath}
\left(T_{g}\right)^{a}_{b}\left(t^{-1}_{g}(x)\right)\psi^{b}\left(t^{-1}_{g}(x)\right)
=\psi^{a}(x),
\end{displaymath}
or equivalently
\begin{equation}\label{translaw}
\left(T_{g}\right)^{a}_{b}(x)\psi^{b}(x)=\psi^{a}\left(t_{g}(x)\right).
\end{equation}
In general a description of all \emph{G}-invariant sections is
hardly possible but under some natural requirements imposed on the
group action it may be achieved. There is a "smaller" reduced
bundle $\xi_G$ that smooth sections (without any restriction) in
it are one-to-one correspondent with $G$-invariant sections in
$\xi$ (elements of $C^{\infty}(\xi)_{G}$). The abstract algebraic
construction of $\xi_G$ consists of two steps. Consider the
stationary subgroup $H_{x_0}=\{h\in G|t_h(x_0)=x_0\}$ of a point
$x_0\in M$. For $x=x_0$ the $G$-invariance condition
(\ref{translaw}) is
\begin{equation}
\left(T_{h}\right)^{a}_{b}(x_0)\psi^{b}(x_0)=\psi^{a}(x_0)\; ,\quad
h\in H_{x_0}.
\end{equation}
As a matter of fact, this is a restriction on the values of
$G$-invariant sections at any given point $x_0$. Let
$st(\xi)_{x_0}=\{\overrightarrow{u}\in\xi_{x_0}|T_h(\overrightarrow{u})
=\overrightarrow{u},\forall\,h\in H_{x_0}\}$ be the vector
subspace of all fixed vectors. We assume that the collection of
all spaces $st(\xi)_x$ for all $x\in M$ is a vector bundle
$st(\xi)\subset\xi$ called a stationary subbundle. This condition
limits the action of $G$. Obviously, $C^{\infty}(\xi)_G\subset
C^{\infty}(st(\xi))$.

\textbf{Note}: The explicit construction of the stationary
subbundle $st(\xi)$ is the crucial point where the new structure
of the reduced bundle arises. This is the most difficult step in
our approach.

The second step consists in taking the quotient of the base
\emph{M} for the bundle $st(\xi)$. We suppose that the projection
$t_g$ has uniform orbits in the base \emph{M} and \emph{M} itself
represents a total space of a smooth locally trivial bundle
$(M,\pi,M/G)$ of homogeneous spaces. It is another requirement on
the action of the group $G$. Let's consider an orbit of $G$
linking up $x,y\in M$, i. e. $\exists\,g\in G:\:y=t_g(x)$. If
$\psi\in C^{\infty}(\xi)_G$ the value $\psi(y)$ is uniquely
determined by the value of the section at $x$ (in accordance to
the invariance condition (\ref{translaw}))
\begin{equation}\label{invcon}
\psi^a(y)=\psi^a(t_g(x))=(T_g)^a_b(x)\psi^b(x).
\end{equation}

Let $(M,\pi,M/G)$ be a trivial bundle and \emph{N} be a global
section i.e. $N\subset M$ is transversal to the orbits. Because of
the relation ($\ref{invcon}$), a \emph{G}-invariant section $\psi$
is completely determined if we know the restriction
$\psi|_N$($\psi|_N\in C^{\infty}(st(\xi)|_N)$). Moreover, if
$\varphi\in C^{\infty}(st(\xi)|_N)$ then $\varphi$ correctly
induces an invariant section $\psi\in C^{\infty}(\xi)_G$. Indeed,
let $y\in M$, then the orbit through $y$ intersects the
submanifold $N$ only in one point $x$ : $ y=t_g(x)$ for some $g\in
G$. By definition, the induced \emph{G}-invariant section is
\begin{displaymath}
\psi^a(y):=\left(T_g\right)^a_b(x)\varphi^b(x).
\end{displaymath}

The element $g\in G$ is not uniquely determined by $y$ and $x$ but
since $\varphi(x)\in st(\xi)_x$ the value $\psi(y)$ doesn't depend
on the specific choice of the group element. The restriction
$st(\xi)|_N$ is a coordinate realization of the bundle $\xi_G$. If
we consider another submanifold $N'\subset M$ transversal to the
orbits in $M$, the corresponding restriction $st(\xi)|_{N'}$ is
another coordinate realization. There is a canonical isomorphism
$st(\xi)_{N'}\approx st(\xi)){N}$ induced by the group action.
This procedure sews the abstract reduced bundle from the
coordinate realization. If $(M,\pi,M/G)$ isn't trivial the
construction of $\xi_G$ is analogous but slightly complicated. We
have to sew local coordinate realizations. Smooth sections in the
reduced bundle are one-to-one correspondent with
\emph{G}-invariant sections. We shall transform the problem of
describing invariant differential operators into a problem for
characterization of invariant sections in appropriate jet bundles.

Consider two bundles $\xi$ and $\eta$ over the same base \emph{M}.
A group \emph{G} acts on both of them by the same projection
\emph{t} in \emph{M}. The action of $G$ on $C^{\infty}(\xi)$ and
$C^{\infty}(\eta)$ induces an action on differential operators
$D:C^{\infty}(\xi)\to C^{\infty}(\eta)$
\begin{equation}
g(D)(\psi):=g\left(D\left(g^{-1}(\psi)\right)\right).
\end{equation}
We use for simplicity the same notation for the actions of
\emph{G} on $\xi$ and on $\eta$ . A differential operator is
called \emph{G}-invariant if it satisfies the following condition
\begin{equation}
g(D)(\psi)=D\left(g(\psi)\right),\quad\forall\,g\in G,\;
\forall\,\psi\in C^{\infty}(\xi).
\end{equation}
The problem we consider is the description of all invariant
differential operators. This problem can be reduced to the problem
we studied before by using the jet bundle technique (for more
details about jet bundles see \cite{jet1},\cite{jet2},\cite{jet3},
\cite{jet4}). Let a differential operator $D:C^{\infty}(\xi)\to
C^{\infty}(\eta)$ be of order (up to) \emph{k} and linear. We
denote with $J^{k}(\xi)$ the corresponding $k$-jet bundle of
$\xi$. For a local coordinate frame $(x^{\mu},u^{a})$ in $\xi$
there exists an induced coordinate frame in $J^k(\xi)$ denoted by
$(x^{\mu},u^{a},u^{a}_{\mu},\ldots,u^{a}_{\mu_1\mu_2\ldots\mu_k})$,
where the indices are ordered $\mu_1\leq\mu_2\leq\ldots\leq\mu_k$.
Each vector $u\in J^k(\xi)_x$ is a jet of particular section
$\psi:M\to E$, i.e. $u=j^k(\psi)_x$ so that
\begin{displaymath}
u^a=\psi^a(x),\quad u^a_{\mu}=\partial_{\mu}\psi^a(x),\ldots,\quad
u^a_{\mu_1\mu_2\ldots\mu_k}=\partial_{\mu_1\mu_2\ldots\mu_k}\psi^a(x).
\end{displaymath}
Any linear differential operator is completely determined by its
general symbol, i.e. the bundle morphism
$\mathscr{D}:J^k(\xi)\to\eta$ (over the identity on \emph{M}).
Using some natural isomorphisms, one can view general symbols as
sections in the tensor product
$\left(J^k(\xi)\right)^{\ast}\otimes\,\eta$. The set of all linear
differential operators of order up to $k$ corresponds one-to-one
to smooth sections
$C^{\infty}\left(\left(J^k(\xi)\right)^{\ast}\otimes\,\eta\right)$.
The jet lifting of the sections $j^k:C^{\infty}(\xi)\to
C^{\infty}\left(J^k(\xi)\right)$ plays the role of an universal
differential operator of order (up to) $k$. An arbitrary linear
differential operator $D:C^{\infty}(\xi)\to C^{\infty}(\eta)$ of
order $k$ is a composition $D=\mathscr{D}\circ j^k$. If $D$ is a
nonlinear operator then its general symbol is a fibre-preserving
map $\mathscr{D}:J^k(\xi)\to\eta$. The action of the group $G$ in
the bundle $\xi$ induces another action of $G$ in the
corresponding jet bundle $J^k(\xi)$(so-called jet lifting of the
action). Thus we have an action of \emph{G} in the tensor product
$\left(J^k(\xi)\right)^{\ast}\otimes\,\eta$. Then a differential
operator is invariant if and only if its general symbol is an
invariant section in $\left(J^k(\xi)\right)^{\ast}\otimes\,\eta$.
So the jet bundle technique gives an algebraization of the
(linear) differential operators. If we construct the reduced
bundle of $\left(J^k(\xi)\right)^{\ast}\otimes\,\eta$ this will
provide a full description of all invariant operators. In the
nonlinear case general symbol $mathscr{D}$ is a fibre-preserving
map. At any point $x\in M$ the restriction
$\mathscr{D}_x:\left(J^k(\xi)_x\to\eta_x\right)$ is a smooth
(nonlinear) map. The description of the \emph{G}-invariant
nonlinear operators is similar to the linear case. We study the
action of the stationary subgroup $H_{x_0}$ on the space
$C^{\infty}\left(\left(J^k(\xi)\right)_{x_0},\eta_{x_0}\right)$----
the space of all smooth maps $J^k(\xi)_{x_0}\to\eta_{x_0}$ and
then we have to find the fixed elements in it. We use this scheme
to describe all conformally invariant differential operators
acting on Minkowski space. As a simple illustration we will
consider the two dimensional algebraic conformal case. In both
cases the base \emph{M} is a homogeneous space. The reduced bundle
$\xi_G$ consists of one fibre over one point. The crucial point is
to find the stationary elements of $C^{\infty}(\xi)_{x_0}$ for
only one point.

\section{Illustrative example}

We consider the space of analytic functions of a single complex
variable and the differential operators between analytic
functions. The analytic functions may be viewed as sections on
trivial line bundles over the complex plane $\mathbb{C}$, i.e.
$\xi=(\mathbb{C}\times\mathbb{C},p,\mathbb{C})$. The group
$GL(2,\mathbb{C})$ acts on $\mathbb{C}$ by rational
transformations
\begin{equation}
t_g(z):=\frac{az+b}{cz+d},\quad z\in\mathbb{C},\,\quad
g=\left(\begin{array}{cc} a & b\\ c & d\end{array}\right)\in
GL(2,\mathbb{C})
\end{equation}
$GL(2,\mathbb{C})$ acts on the analytic functions by transforming
the argument. In adapted coordinates $(z,u)$ this action is
\begin{equation}
g(z,u)=\left(\frac{az+b}{cz+d},u\right).
\end{equation}
The problem to solve is the description of all
$GL(2,\mathbb{C})$-invariant differential operators. The
invariance condition now reads
\begin{displaymath}
D\left(f\left(\frac{az+b}{cz+d}\right)\right)
=\left(D(f)\right)\left(\frac{az+b}{cz+d}\right),\forall\quad
g=\left(\begin{array}{cc} a & b\\ c & d\end{array}\right)\in
GL(2,\mathbb{C}).
\end{displaymath}
The complex plane is a homogeneous space, because translations in
$\mathbb{C}$ are a subgroup of $GL\left(2,\mathbb{C}\right)$
\begin{displaymath}
g=\left(\begin{array}{cc} 1 & b \\ 0 & 1
\end{array}\right)\quad\Longrightarrow\quad t_g(z)=z+b.
\end{displaymath}
We choose $z_0=0$. Its stationary subgroup $H_0$ is defined as the
subset
\begin{equation}
H_{0}=\left\{h\in GL(2,\mathbb{C})\left|\,h=\left(\begin{array}{cc}
a & 0
\\ c & 1 \end{array}\right)\right.\quad a\neq 0\right\}.
\end{equation}

The fibre $J^k(\xi)_0$ is the set of all $k$-jets taken at $z=0$
of analytic functions. The $k$-jet of an analytic function is its
Taylor expansion up to order $k$. If $w=j^k(f)_0\in J^{k}(\xi)_0$
then
\begin{equation}
w=\sum_{l=1}^{k}{1\over l!}u_lz^l.
\end{equation}
where
\[u_l=\frac{d^lf(0)}{dz^l},\quad l=1,2,\ldots,k .\]
We have assumed above for simplicity that $u_0=0$ since it doesn't
lead to any loss of generality.

The prolonged action of $H_0$ on the fibre $J^k(\xi)_0$ is given
by the definition
\begin{equation}\label{barover}
h(w)=j^k\left(f\left(\frac{az}{cz+1}\right)\right)=\overline{j^k(f)\circ
j^k\left(\frac{az}{cz+1}\right)},\quad h\in H_0.
\end{equation}
The bar over the right hand side of (\ref{barover}) indicates that
the composition is truncated, i.e. all monomials of higher order
than $k$ have been ignored. The jet of a composition of functions
is the composition of their jets. On the other hand, the truncated
Taylor expansion has the form
\begin{equation}
j^k\left(\frac{az}{cz+1}\right)=az\sum_{m=0}^{k-1}(-cz)^m.
\end{equation}

The transformation $h(u_1,\ldots,u_k)=(w_1,\ldots,w_k)$ is
determined by the equation
\begin{equation}
\overline{\sum_{l=1}^k{1\over
l!}u_l\left(az\sum_{m=0}^{k-1}(-cz)^m\right)^l}=\sum_{l=1}^k{1\over
l!}w_lz^l
\end{equation}
For example, the transformation of the jet of fourth order is
\begin{equation}
\left|\begin{array}{ccl}
w_1 & = & u_1a \\
w_2 & = & u_2a^2-2u_1ac \\
w_3 & = & u_3a^3-6u_2a^2c+6u_1ac^2 \\
w_4 & = & u_4a^4-12u_3a^3c+36u_2a^2c^2-24u_1ac^3.
\end{array}\right.
\end{equation}

Since we consider only scalar functions we have to describe maps
$J^k(\xi)_0\to\mathbb{C}$ invariant under the action of $H_0$. Let
$J^k(\xi)_0/H_0$ be the quotient space and $\pi :J^k(\xi)_0\to
J^k(\xi)_0/H_0$ be the canonical projection on it. A map
$\mathscr{D}_0:J^k(\xi)_0\to\mathbb{C}$ is said to be
$H_0$-invariant if and only if there exists another map
$\tilde{\mathscr{D}}_0:J^k(\xi)_0/H_0\to\mathbb{C}$ such that the
relation $\mathscr{D}=\tilde{\mathscr{D}}\circ\pi$ holds. One may
rewrite that requirement in terms of commutative diagrams as
follows
\begin{equation}
\begin{array}{ccccc}
            & & \pi & & \\
J^k(\xi)_0 & & \longrightarrow & & J^k(\xi)_0/H_0 \\
            & &                 & & \\
\mathscr{D} & \searrow & & \swarrow & \tilde{\mathscr{D}}\quad \\
            & &                 & & \\
            & & \mathbb{C}      & &
\end{array}.
\end{equation}
In this sense the canonical projection $\pi$ is a universal
$H_0$-invariant map. The components of $\pi$ (in any coordinates
in the quotient space) are invariant. Any $H_0$-invariant map is a
function of the components of $\pi$.

To describe the quotient space means we must find a canonical
representative in every orbit of $H_0$. The jets with $u_1=0$ is an
invariant subspace. Considering the general case with $u_1\neq 0$ in
each orbit there is an unique representative with $u_1=1$ and
$u_2=0$. The projection on this canonical representative is given by
the element $h\in H_0$ with $a=1/u_1,\, c=u_2/2u_1$ more precisely
\begin{equation}\label{genpol}
\overline{\sum_{l=1}^k{1\over l!}u_l\left({1\over
u_1}z\sum_{m=0}^{k-1}\left(-\frac{u_2}{2u_1}z\right)^m\right)^l}=
z+\sum_{l=3}^k{1\over l!}w_lz^l.
\end{equation}
The coefficients $w_l$ are coordinates in the quotient space
\begin{equation}\label{invarsym1}
w_3=\frac{u_3}{\left(u_1\right)^3}-{3\over
2}\left(\frac{u_2}{\left(u_1\right)^2}\right)^2
\end{equation}

\begin{equation}\label{invarsym2}
w_4=\frac{u_4}{\left(u_1\right)^4}-6\frac{u_2u_3}{\left(u_1\right)^5}
+6\frac{\left(u_2\right)^3}{\left(u_1\right)^6},
\end{equation}

\begin{equation}w_5=\frac{u_5}{\left(u_1\right)^5}-10\frac{u_2u_4}{\left(u_1\right)^6}+
30\frac{\left( u_2\right)^2u_3}{\left(
u_1\right)^7}-\frac{45}{2}\frac{\left(u_2\right)^4}{\left(u_1\right)^8}
\end{equation}

The coefficients $w_l$ are the general symbols of the
$H_0$-invariant differential operators at $z=0$. Since
translations act in a trivial manner on the $k$-jets they look
exactly the same at any point of $\mathbb{C}$. If we consider the
infinite order jets by using the previous method we can obtain the
following generating power series
\begin{equation}
\sum_{k=1}^{\infty}{1\over k!}u_k\left({z\over
u_1}\sum_{l=1}^{\infty}\left(-\frac{u_2}{2u_1}z\right)^l\right)^k
=z+\sum_{k=3}^{\infty}{1\over k!}w_kz^k.
\end{equation}
The functions $w_l=w_l(u_1,u_2,\ldots,u_l),\quad l=3,4,\ldots$ are
components of the canonical projection $\pi$. According to the
previous remark the corresponding differential operators are
universal (a complete system of invariants). Any invariant
differential operator up to order $k$ is a function of them. The
invariant differential operators corresponding to the symbols
(\ref{invarsym1}) and (\ref{invarsym2}) look as follows
\begin{equation}\label{Schwarz}
D_1(f)={1\over
\left(f'(z)\right)^2}\left(\frac{f'''(z)}{f'(z)}-{3\over
2}\left(\frac{f''(z)}{f'(z)}\right)^2\right),
\end{equation}
and
\begin{displaymath}
D_2(f)=\frac{f^{(IV)}(z)}{\left(f'(z)\right)^4}
-6\frac{f'''(z)f''(z)}{\left(f'(z)\right)^5}
+6\frac{\left(f''(z)\right)^3}{\left(f'(z)\right)^6}.
\end{displaymath}
The expression in the parenthesis in (\ref{Schwarz}) is the
well-known Schwarz derivative. Thus we have confirmed the classic
result that the Schwarz derivative is an invariant differential
operator with respect to the action of fraction-linear
transformations. The Schwarz derivative takes values in quadratic
differentials. The difference between (\ref{Schwarz}) and the
Schwarz derivative is the factor $\left(f'(z)\right)^{-2}$ because
$D_1(f)$ takes values in scalar functions. Further more, we have
obtained that there are $k-2$ functionally independent
$GL(2,\mathbb{C})$-invariant differential operators of order
$l\leq k$. Any invariant differential operator of the same order
is a function of them.

Each invariant differential operator is nonlinear and it has a
particular domain, i.e. it exists the requirement $f'(z)\neq 0$.

\textbf{Note}: The fact that the $k$-jet $j^k(f)_0$ starts with 0
enables us to exclude the trivial case when the differential
operators are operators of zero order, i.e. functions of $f$. It is
clear that such operators are for sure $GL(2,\mathbb{C})$-invariant.
Since we exclude the invariant subspace of jets with $u_1=0$ we
obtain in general differential operators which involve division by a
power of $f'(z)$.

\section{Minkowski case}

Let us consider the Minkowski space
$\mathscr{M}=\left(\mathbb{R}^4,\eta\right)$ with the
pseudoeuclidean metric tensor $\eta=diag(-1,1,1,1)$. The conformal
group $C(1,3)$ consists of all diffeomorphisms $\varphi
:\mathbb{R}^4\to\mathbb{R}^4$ preserving the conformal class of
$\eta$, i.e.
\begin{displaymath}
\varphi^{\ast}(\eta)_{\mu\nu}(x)=\frac{\partial\varphi^{\alpha}}{\partial
x^{\mu}}(x)\frac{\partial\varphi^{\beta}}{\partial
x^{\nu}}(x)\eta_{\alpha\beta}(\varphi(x))=e^{w(x)}\eta_{\mu\nu}(x),
\end{displaymath}
where $w(x)$ is a smooth function( the Greek indices take the
values $0,1,2,3$). The conformal group $C(1,3)$ is induced by the
following transformations
\begin{enumerate}
    \item translations: $x^{\mu}\to x^{\mu}+a^{\mu}$,
    \quad $a\in\mathbb{R}^{4}$;
    \item rotations: $x^{\mu}\to\Lambda^{\mu}_{\nu}x^{\nu}$,
    \quad $\Lambda^{\mu}_{\alpha}\eta_{\mu\nu}\Lambda^{\nu}_{\beta}=
    \eta_{\alpha\beta}$;
    \item dilatation: $x^{\mu}\to\lambda x^{\mu}$,
    \quad $\lambda>0$;
    \item special conformal transformations:
    $x^{\mu}\to\left(x^{\mu}+b^{\mu}x^{2}\right)
    /\left(1+2x.b+
    x^2b^{2}\right),\quad b\in \mathbb{R}^4
    $.
\end{enumerate}
\textbf{Note}: The special conformal transformations have a correct
global definition in the compactified Minkowski space.

The conformal group acts on the space of the smooth scalar
functions $\mathbb{R}^4\to \mathbb{R}$ by transforming the
argument. We are looking for differential operator between scalar
functions invariant under the action of the conformal group. We
are going to follow the scheme demonstrated in the precedent
section.

The Minkowski space is a homogeneous space. The stationary
subgroup $H_0\subset C(1,3)$ for $x=0$ is induced by the
transformations
\begin{enumerate}
    \item $x^{\mu}\to\Lambda^{\mu}_{\nu}x^{\nu}$ ;
    \item $x^{\mu}\to\lambda x^{\mu}$ ;
    \item $x^{\mu}\to\left(x^{\mu}+b^{\mu}x^{2}\right)/
    \left(1+2x.b
    +x^2b^2\right)$.
\end{enumerate}
We must consider the action of $H_0$ on the jets
$J^k\left(\mathbb{R}^4\right)_0=\left\{j^k(f)_0\;|\; f\in
C^{\infty}\left(\mathbb{R}\right)^4,\; f(0)=0\right\}$ and find the
$H_0$-invariant functions
$J^k\left(\mathbb{R}^4\right)_0\to\mathbb{R}$, i.e. we have to
describe the canonical projection $\pi:J^k\left(\mathbb{R}^4
\right)_0\to J^k\left(\mathbb{R}^4\right)_0/H_0$.

The first nontrivial case is $k=2$. The 2-jet of a smooth function
$f$ is the Taylor polynimial
\begin{equation}
j^2(f)_0=u_{\alpha}x^{\alpha}+{1\over
2}u_{\alpha_1\alpha_2}x^{\alpha_1}x^{\alpha_2}.
\end{equation}
where $(u_{\alpha},u_{\alpha_{1}\alpha_2}),\,
\alpha_1\leq\alpha_2$ represent coordinates in the fibre
$J^k\left(\mathbb{R}^4\right)_0$.

Let $\varphi :\mathbb{R}^{4}\to\mathbb{R}^{4}$ be a diffeomorphism
with a fixed point $x=0$ then its second order jet at $x=0$ is
\begin{equation}
\left(j^{2}(\varphi)_{0}\right)^{\mu}= A_{\nu}^{\mu}x^{\nu}+{1\over
2}A^{\mu}_{\nu_{1}\nu_{2}}x^{\nu_{1}}x^{\nu_{2}},
\end{equation}
assuming that $det\left(A^{\mu}_{\nu}\right)\neq 0$
($A^{\mu}_{\nu_1\nu_2}$ is arbitrary). The action of a
diffeomorphism $\varphi$ on the space
$J^2\left(\mathbb{R}^4\right)_0$ is given by
\begin{eqnarray}\label{diffactjet2}
\left(\begin{array}{c}u_{\mu}\\u_{\mu_{1}\mu_{2}}
\end{array}\right) & \to & \left(\begin{array}{c}
A_{\mu}^{\nu}u_{\nu}\\
A_{\mu_{1}}^{\nu_{1}}A_{\mu_{2}}^{\nu_{2}}
u_{\nu_{1}\nu_{2}}+A_{\mu_{1}\mu_{2}}^{\nu}u_{\nu}
\end{array}\right).
\end{eqnarray}
A special case is the action of the stationary subgroup $H_0$. We
have for the prolonged action of the dilatation
\begin{eqnarray}\label{dilactjet2}
\left(\begin{array}{c} u_{\mu}\\ u_{\mu_{1}\mu_{2}}
\end{array}\right) & \to & \left(\begin{array}{c}
\lambda u_{\mu}\\
\lambda^{2} u_{\mu_{1}\mu_{2}}\end{array}\right)
\end{eqnarray}
 as well as for rotations
\begin{eqnarray}
\left(\begin{array}{c} u_{\mu}\\ u_{\mu_{1}\mu_{2}}
\end{array}\right) & \to & \left(\begin{array}{c}
\Lambda_{\mu}^{\nu}u_{\nu}\\
\Lambda_{\mu_{1}}^{\nu_{1}}\Lambda_{\mu_{2}}^{\nu_{2}}
u_{\nu_{1}\nu_{2}}\end{array}\right).
\end{eqnarray}
The infinite jet of special conformal transformations $\varphi$
reads
\[\left(j\,^{\infty}(\varphi(x))|_0\right)^{\mu}
=\left(x^{\mu}+b^{\mu}x^2\right)
\sum_{k=0}^{\infty}(-1)^k\left(2b. x+b^2x^2\right)^k,\]
specifically for $k=2$
\begin{equation}\label{difjet2}
\left(j\,^2(\varphi(x))|_0\right)^{\mu}=x^{\mu}-
2x^{\mu}b.x+b^{\mu}x^2.
\end{equation}
Thus the action is
\begin{eqnarray}\label{invactjet2}
\left(\begin{array}{c} u_{\mu}\\ u_{\mu_{1}\mu_{2}}
\end{array}\right) & \to & \left(\begin{array}{c} u_{\mu}\\
u_{\mu_{1}\mu_{2}}-2u_{\mu_{1}}b_{\mu_{2}} -2u_{\mu_{2}}b_{\mu_{1}}
+2u_{\beta}b^{\beta}\eta_{\mu_{1}\mu_{2}}\end{array}\right),
\end{eqnarray}
where $b_{\mu}=\eta_{\mu\nu}b^{\nu}$.

We will describe the quotient space
$J^2\left(\mathbb{R}^4\right)_0/H_0$  by choosing a canonical
representative from each orbit of $H_0$. Let
$(u_{\alpha},u_{\alpha_1\alpha_2})$ be coordinates in
$J^2\left(\mathbb{R}^4\right)_0$.

At this step we assume that $u^2:=\eta_{\mu\nu}u^{\mu}u^{\nu}<0$. By
dilatation choosing $\lambda=1/\sqrt{\left(-u^2\right)}$ we obtain
another representative of the same equivalence class
\begin{eqnarray}
\left(\begin{array}{c} u_{\alpha}\\ u_{\alpha_{1}\alpha_{2}}
\end{array}\right) & \to & \left(\begin{array}{c}v_{\alpha}\\
v_{\alpha_{1}\alpha_{2}} \end{array}\right)=\left(\begin{array}{c}
u_{\alpha}/ \sqrt{-u^2}\\
u_{\alpha_{1}\alpha_{2}}/ \left(-u^2\right)
\end{array}\right).
\end{eqnarray}
By a special hyperbolic rotation
\begin{equation}\label{boost}
A^{\beta}_{\alpha}=\left(\begin{array}{cc} v_0 & -\overrightarrow{v} \\
\overrightarrow{v} &
\delta^i_j-v^iv_j\left(1+v_0\right)/\overrightarrow{v}^2
\end{array}\right),
\end{equation}
where we denote by $\overrightarrow{v}$ the space component
$\left(v^1,v^2,v^3\right)$ of the 4-vector $v$, we reach to another
representative
\begin{eqnarray}\label{simljet2}
\left(\begin{array}{c} v_{\alpha} \\
v_{\alpha_{1}\alpha_{2}} \end{array}\right) & \to &
\left(\begin{array}{c}(1,0,0,0)\\
A_{\alpha_1}^{\beta_1}A_{\alpha_2}^{\beta_2}v_{\beta_1\beta_2}
\end{array}\right)=\left(\begin{array}{c}e_0\\
w_{\alpha_1\alpha_2}\end{array}\right).
\end{eqnarray}
In the case of 2-jets looking like (\ref{simljet2}) the action
(\ref{invactjet2}) reads
\begin{displaymath}
\left|\begin{array}{ccl}e_0 & \to & e_0 \\
w_{0\beta} & \to & w_{0\beta}-2a_{\beta}\\
w_{ij} & \to & w_{ij}-2a_0\delta_{ij}.\end{array}\right.
\end{displaymath}
This formula enables us to choose $a_{\beta}=w_{0\beta}/2$ and
therefore we get another "more canonical" representative
\begin{eqnarray}\label{canon}
\left(\begin{array}{c} e_0 \\
\left(w_{\mu\nu}\right)
\end{array}\right) & \to & \left(\begin{array}{c}(1,0,0,0)\\
\left(\begin{array}{cc}0 & \overrightarrow{0}\\
\overrightarrow{0} & \tilde{w}_{mn}\end{array}\right)
\end{array}\right)=:\left(\begin{array}{c}e_0 \\
w_{\alpha_1\alpha_2}\end{array}\right).
\end{eqnarray}
Each equivalence class of this type of jets has got such a
representative. The subgroup $O(3)\subset C(1,3)$ is still acting
on the jets (\ref{canon}) preserving their form. The action on
symmetric $3\times 3$ matrices is $\tilde{w}\to B\tilde{w}B^{T}$,
$B\in O(3)$. That is why we can choose the transformation (a
different boost) turning the matrix $\tilde{w}$ into a diagonal
form
\begin{eqnarray*}
\left(\begin{array}{c} (1,0,0,0)\\ \left(\begin{array}{cc} 0 &
\overrightarrow{0} \\ \overrightarrow{0} &
\tilde{w}_{mn}\end{array}\right)
\end{array}\right) & \to & \left(\begin{array}{c} (1,0,0,0)\\
\left(\begin{array}{cc} 0 & \overrightarrow{0}\\
\overrightarrow{0} &
diag\left(\lambda_{1},\lambda_{2},\lambda_{3}\right)\end{array}\right)
\end{array}\right).
\end{eqnarray*}
The unordered triple of eigenvalues $(\lambda_1,
\lambda_2,\lambda_3)$ are coordinates in the quotient space
$J^2(\mathbb{R}^4)_0/H_0$, so they describe completely this
quotient space. As coordinate frame we may choose the elementary
symmetric polynomials
\[\sigma_1=\lambda_1+\lambda_2+\lambda_3,\quad \sigma_2=\lambda_1\lambda_2
+\lambda_1\lambda_3+\lambda_2\lambda_3,\quad
\sigma_3=\lambda_1\lambda_2\lambda_3.
\]
but we prefer working with the following frame
\begin{eqnarray}\label{powers}
S_1\left(\lambda_1,\lambda_2,\lambda_3\right) & = &
\lambda_{1}+\lambda_{2}+\lambda_{3}\nonumber\\
S_2\left(\lambda_1,\lambda_2,\lambda_3\right) & = &
\left(\lambda_1\right)^2+\left(\lambda_2\right)^2+\left(\lambda_3\right)^2\\
S_3\left(\lambda_1,\lambda_2,\lambda_3\right) & = &
\left(\lambda_1\right)^3+\left(\lambda_2\right)^3+\left(\lambda_3\right)^3.\nonumber
\end{eqnarray}
because of its convenient form
\begin{equation}\label{traces}
S_k(\lambda_1,\lambda_2,\lambda_3)=Tr(\tilde{w}^k),\quad k=1,2,3.
\end{equation}
The quantities $S_1$,$S_2$ and $S_3$ are coordinates in the
equivalence class of the jet that we started from. To obtain them
as explicit functions of the initial jet we must take the
composition
\begin{displaymath}
\left(\begin{array}{c}u_{\alpha}\\
u_{\alpha_1\alpha_2}\end{array}\right) \to \left(\begin{array}{c}
v_{\alpha}\\ v_{\alpha_1\alpha_2}\end{array}\right)\to
\left(\begin{array}{c}e_0 \\
w_{\alpha_1\alpha_2}\end{array}\right)\to
\left(\begin{array}{c}e_0 \\
\tilde{w}_{\alpha_1\alpha_2}\end{array}\right)\to
\left(\begin{array}{c}Tr(\tilde{w})\\Tr(\tilde{w})^2\\Tr(\tilde{w})^3
\end{array}\right).
\end{displaymath}
The functions $\mathscr{D}_k=Tr\left(\tilde{w}^k\right)=
\mathscr{D}_k\left(u_{\alpha},u_{\alpha_1\alpha_2}\right)$ are the
general symbols of the invariant operators at $x=0$. Since the
translations act trivially on the jets the differential operators
look in the same way at any point of the Minkowski space. The
final result is
\begin{equation}\label{D1}
D_{1}(f)=\frac{\nabla^{2} f}{(\nabla
f)^{2}}+2\frac{\partial^{\alpha}f\ \partial^{\beta}f} {(\nabla
f)^2}\frac{\partial_{\alpha\beta} f}{(\nabla f)^2}.
\end{equation}

\begin{eqnarray}\label{D2}
D_{2}(f)& = &
\left(\eta^{\alpha\mu}\eta^{\beta\lambda}-2\eta^{\alpha\mu}\frac{\partial^{\beta}
f\ \partial^{\lambda}f}{(\nabla f)^{2}}+2\frac{\partial^{\alpha}
f\ \partial^{\beta}f}{(\nabla f)^{2}}\frac{\partial^{\lambda} f\
\partial^{\mu}f}{(\nabla f)^{2}}\right)
\frac{\partial_{\alpha\beta} f}{(\nabla
f)^{2}}\frac{\partial_{\lambda\mu} f}{(\nabla f)^{2}}+\nonumber\\
& & 2\frac{\nabla^{2}f}{(\nabla f)^{2}}\frac{\partial^{\alpha} f\
\partial^{\beta} f}{(\nabla f)^{2}}\frac{\partial_{\alpha\beta}
f}{(\nabla f)^{2}}
\end{eqnarray}

\begin{equation}\label{D3}
D_{3}(f)=
C^{\alpha\beta\lambda\mu\rho\sigma}\frac{\partial_{\alpha\beta}f}{(\nabla
f)^2}\frac{\partial_{\lambda\mu}f}{(\nabla
f)^2}\frac{\partial_{\rho\sigma}f}{(\nabla f)^2}+3\frac{\nabla^2
f}{(\nabla f)^2}\frac{\partial^{\alpha}f\partial^{\beta}f}{(\nabla
f)^2}\frac{\partial^{\lambda}f\partial^{\mu}f}{(\nabla
f)^2}\frac{\partial_{\alpha\beta}f}{(\nabla
f)^2}\frac{\partial_{\lambda\mu}f}{(\nabla f)^2},
\end{equation}
where
\begin{eqnarray*}
C^{\alpha\beta\lambda\mu\rho\sigma}&:=&
\eta^{\alpha\sigma}\eta^{\beta\lambda}\eta^{\mu\rho}-3\eta^{\alpha\sigma}
\eta^{\beta\lambda}\frac{\partial^{\mu}f\partial^{\rho}f}{(\nabla
f)^2}+3\eta^{\alpha\mu}\eta^{\beta\lambda}
\frac{\partial^{\rho}f\partial^{\sigma}f}{(\nabla
f)^2}+\\
& &
3\eta^{\alpha\sigma}\frac{\partial^{\beta}f\partial^{\lambda}f}{(\nabla
f)^2}\frac{\partial^{\mu}f\partial^{\rho}f}{(\nabla
f)^2}-6\eta^{\alpha\mu}\frac{\partial^{\beta}f\partial^{\lambda}f}{(\nabla
f)^2}\frac{\partial^{\rho}f\partial^{\sigma}f}{(\nabla
f)^2}+\\
& & 2\frac{\partial^{\alpha}f\partial^{\beta}f}{(\nabla
f)^2}\frac{\partial^{\lambda}f\partial^{\mu}f}{(\nabla
f)^2}\frac{\partial^{\rho}f\partial^{\sigma}f}{(\nabla f)^2},
\end{eqnarray*}
\[\left(\nabla f\right)^2=\eta^{\alpha\beta}\partial_{\alpha}f\partial_{\beta}f,
\quad \nabla^2 f=\eta^{\alpha\beta}\partial_{\alpha\beta}f\]

So we have just obtained the differential operators having
considered functions with time-like gradients. According to the
general scheme these differential operators are functionally
independent and universal, i.e. they generate all conformally
invariant operators of second order (defined on functions with
time-like gradient). This procedure also contains an algorithm for
calculating the higher order differential operators (defined on
the same subset of functions).

The natural framework of describing the conformally invariant
differential operators involves the complexified and compactified
Minkowski space.

This technique is applicable to the case of $n$-dimensional
pseudoeuclidean space too. There are no conformally invariant first
order differential operators. The conformally invariant differential
operators of second order are generated by $n-1$ functionally
independent differential invariants. For example, one of these
invariant operators (involving functions with time-like gradient) is
the following one
\[D(f)=\frac{\nabla^2f}{\left(\nabla f\right)^2}+(n-2)\frac{\partial^{\alpha}f
\partial^{\beta}f}{\left(\nabla f\right)^2}\frac{\partial_{\alpha\beta}f}
{\left(\nabla f\right)^2}\quad \alpha ,\beta=1\ldots n\].

\textbf{Acknowledgement.}

That study is supported by the contract 35/2003 with Sofia
University.

\end{document}